\RequirePackage{lineno}
\documentclass[prl,twocolumn,amsmath,superscriptaddress]{revtex4}
\usepackage{bm}
\usepackage{amsmath}
\usepackage{amssymb}
\usepackage{amsfonts}
\usepackage{graphicx}
\usepackage{color}

\usepackage{xcolor}
\usepackage{floatrow}
\usepackage{upgreek}
\usepackage{bm}
\DeclareSymbolFont{matha}{OML}{txmi}{m}{it}
\DeclareMathSymbol{\varv}{\mathord}{matha}{118}
\usepackage{natbib}

\begin{document}

\title{Atomic Collapse in Disordered Graphene Quantum Dots}


\author{Mustafa Polat}
\affiliation{Izmir Institute of Technology, Department of Physics, 35430 Urla, Izmir, Turkey}%
\email{mustafapolat@iyte.edu.tr}

\author{A. D. G\"{u}\c{c}l\"{u}}
\affiliation{Izmir Institute of Technology, Department of Physics, 35430 Urla, Izmir, Turkey}%
\date{\today}
\begin{abstract}
In this paper, we numerically study a Coulomb impurity problem
for interacting Dirac fermions restricted in disordered graphene quantum dots. In the presence of
randomly distributed lattice defects and spatial potential fluctuations,
the response of the critical coupling constant for atomic collapse is mainly investigated by local density of states
calculations within the extended mean-field Hubbard model. We find that both types of disorder cause an amplification of the critical threshold. As a result, up to thirty-four percent increase in the critical coupling constant is reported.
This numerical result may explain why the Coulomb impurities remain subcritical in experiments,
even if they are supercritical in theory. Our results also point to the possibility that
atomic collapse can be observed in defect-rich samples such as Ar$^{+}$ ion bombarded, He$^{+}$ ion irradiated,
and hydrogenated graphene.
\end{abstract}

\maketitle

Quantum electrodynamics predicts that the 1S$_{1/2}$ state is only stable up to a critical nuclear charge \(Z_{c}\)
$\sim$ 172; otherwise, formerly bound state becomes a resonant state \cite{reinhardt1977quantum}.
In spite of its long-standing history \cite{greiner2000relativistic}, collapse of the vacuum is far from being proven in experiments performed with real atoms
\cite{cowan1985anomalous}. However, graphene reduces the critical threshold to \(Z_{c}\) $\gtrsim$ 1
through a larger fine structure constant $\alpha$ = 2.2/$\kappa$ \cite{pereira2007coulomb,shytov2007atomic}, where $\kappa$ is the
dielectric constant. Therefore, the idea of creating
an artificial supercritical atom with a smaller critical valence charge has received considerable experimental attention \cite{wang2012mapping,wang2013observing,mao2016realization,wong2017spatially,lu2019frustrated}. In the condensed matter analogue,
Dirac fermions form the vacuum itself, and the Coulomb impurity acts as a
nucleus that couples to the vacuum by means of a dimensionless coupling strength $\beta$ = \(Z\)$\alpha$ \cite{neto2009electronic}.
When $\beta$ exceeds a critical coupling constant $\beta_{c}$, the lowest energy electron state firstly turns into
a quasi-bound state (QBS) \cite{pereira2007coulomb}, which
corresponds to the 1S$_{1/2}$ state of the impurity, and an infinite number of QBS
can appear for massless fermions, depending on the value of $\beta$ \cite{shytov2007atomic}. The critical coupling constant
is estimated to be $\beta_{c}$ = 0.5 for a vacuum consisting of non-interacting massless
Dirac fermions \cite{shytov2007atomic,pereira2007coulomb}, and it remains the same
when these fermions are confined in smaller-sized graphene
quantum dots (GQDs) \cite{van2017graphene,polat2020collapse}.
A further extension of the problem takes electron interactions into account
\cite{biswas2007coulomb,terekhov2008screening} for which this critical value is renormalized to $\beta_{c}$ = 0.6 due to off-site Coulomb repulsion among Dirac particles \cite{polat2020collapse}. However, up until now, all theoretical calculations
assume a disorder-free graphene by ignoring the experimental facts \cite{hashimoto2004direct,martin2008observation},
and the question of effects of imperfections on atomic collapse in graphene has not been addressed yet.

Atomic scale defects \cite{meyer2008direct,banhart2011structural}
and the intercalation of hydrogen atoms \cite{mccreary2012magnetic,wang2018imaging,ccakmak2018effects} may arise during
the growth process, and these defects lead to an imperfect honeycomb lattice \cite{eckmann2012probing,li2019nanoscale}.
Furthermore, such a deformed vacuum can fluctuate in response to spatial charge
inhomogeneities caused by substrate \cite{burson2013direct,ozdemir2016magnetic}.
To find out ambiguous consequences
of these distortions beyond conventional perspective of the theory, the hexagonal GQDs with armchair edges
\cite{gucclu2010excitonic} could provide a
practical playground. These GQDs serve as a bridge between the finite-sized samples
and bulk graphene thanks to their special band-gap characteristics \cite{gucclu2010excitonic,devrim}, and
a sufficiently large size of them is enough to observe atomic collapse, as evidenced by transmission
coefficients of the 1S$_{1/2}$ state \cite{polat2020collapse}. The latter
could help in finding solutions to such complex problems
via exact diagonalization of Hamiltonian, even in the case of
interacting fermions.

In this letter, the critical threshold is studied by placing
the Coulomb impurity at the center of disordered hexagonal
GQDs. Deviations from the perfection in the vacuum are intentionally created by:
(i) randomly distributed point vacancies with different concentrations and (ii) electron-hole puddles
induced by Gaussian impurities. We find a strong dependence of the critical threshold on both types of disorder,
leading to up to thirty-four percent increase in the critical coupling constant.

\begin{figure*}
\floatbox[{\capbeside\thisfloatsetup{capbesideposition={right,center},capbesidewidth=6cm}}]{figure}[\FBwidth]
{\caption{LDOS spectra at the impurity site for : (a) the numbers of 2814, 5514, and 10 806 atoms,
(b) spin-up QBS families, and (c) spin-down QBS families in the presence of finite defect densities. Insets illustrate
zoomed portions of: (a) perfectly ordered and (b) disordered lattices with a central Coulomb impurity. The inset in (c)
is the averaged spin-down DOS that marks the FL at $\tilde{\beta}$ = 0.}\label{fig:fig_1}}
{\includegraphics[width=4.5in]{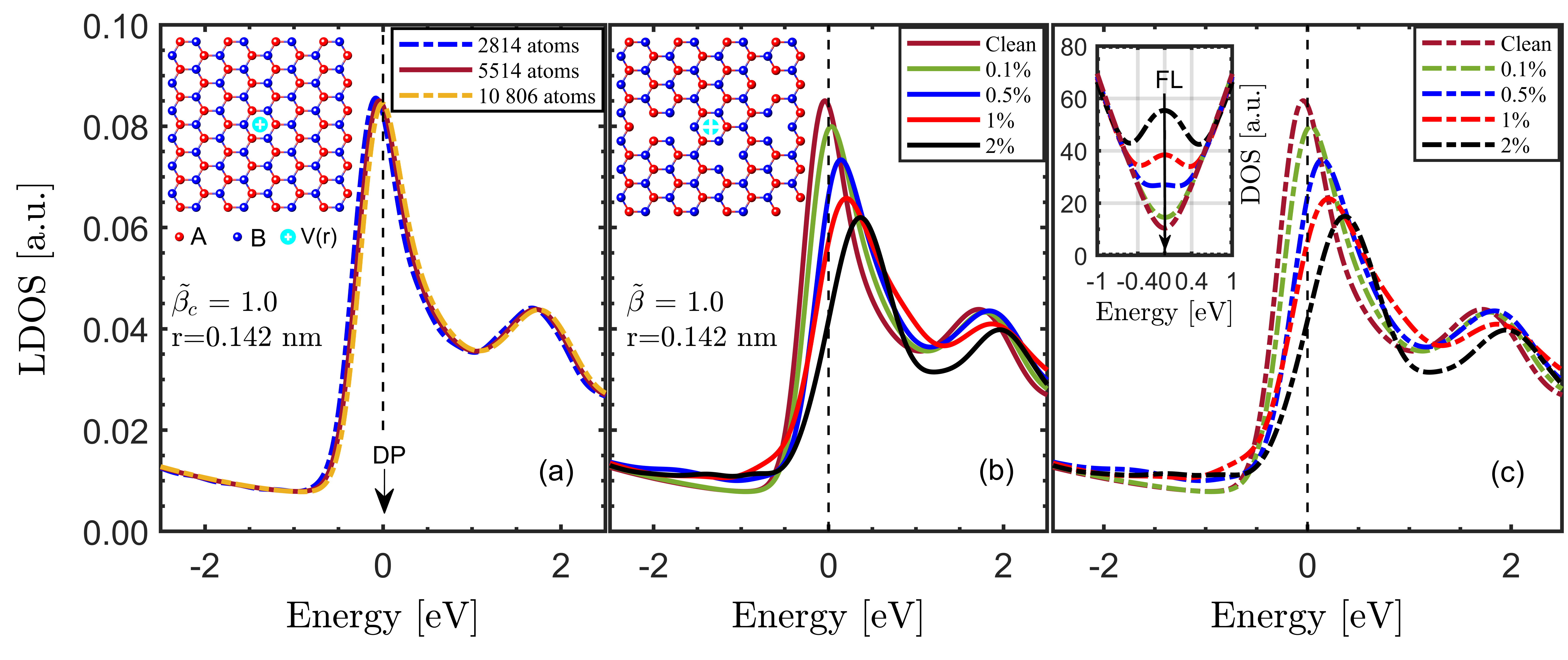}}
\end{figure*}

Extended mean-field Hubbard model is employed
to study the $\pi_{z}$ dynamics, and its Hamiltonian reads
\begin{align}\label{eq:eq1}
H_{MFH} \nonumber & = t\sideset{}{}\sum_{\left<ij\right>\sigma}\bm{\left(}c_{i\sigma}^{\dagger}c_{j\sigma}+\text{H.c.}\bm{\right)} + U\sideset{}{}\sum_{i\sigma}\bm{\left(}\left<n_{i\overline{\sigma}}\right>-\frac{1}{2}\bm{\right)}n_{i\sigma}\\
        & + \sum_{ij}V_{ij}\bm{\left(}\left<n_{j}\right>-1\bm{\right)}n_{i} -\hbar \varv_{\text{F}} \beta\sum_{i\sigma}\frac{c_{i\sigma}^{\dagger}c_{i\sigma}}{r_{i}}\text{,}
\end{align}
where the first term describes the tight-binding Hamiltonian
with a hopping amplitude of \(t\) = -2.8 eV in which the
operator \(c_{i\sigma}^{\dagger}\) (\(c_{i\sigma}\)) creates (annihilates) an electron with
spin $\sigma$ at the lattice site \emph{i}. \(U\) = 16.52/$\kappa$ eV is the onsite
Coulomb repulsion \cite{devrim}, where $\kappa$ = 6 is equivalent to that of
the SiO$_{2}$ substrate under  the effects of interband polarization \cite{ando2006screening}.
\(\left<n_{i\sigma}\right>\) is the spin-dependent expectation value of electron densities,
and n$_{i\sigma}$ is the spin-dependent number operator. Third term \(V_{ij}\) is associated with
the off-site Coulomb repulsion, which is set to be 8.64/$\kappa$
eV, 5.33/$\kappa$ eV, and 27.21/$\kappa$d$_{ij}$ eV for the nearest-neighbors,
next-nearest-neighbors, and the remote atomic sites, respectively
\cite{devrim,potasz2010spin}. d$_{ij}$ is the distance between the
sites \emph{i} and \emph{j} at relatively large distances, and it is in atomic units.
The last term represents the Coulomb impurity placed at the origin of coordinate system, where r$_{i}$ is the
distance between the impurity and the site \emph{i}.
$\varv_{\text{F}}$ $\approx$ 1 $\times$ 10$^{6}$ m/s is the Fermi velocity.

Atomic vacancies with concentrations of 0.1$\%$,
0.5$\%$, 1$\%$, and 2$\%$, which refers to the ratio of the number of point
vacancies \emph{N$_{\text{vac}}$} to that of the lattice sites \emph{N}, are created by
randomly and equally removing the two
sublattices, A (50$\%$) and B (50$\%$), of the
bipartite lattice \cite{altintacs2018defect}.
For 1$\%$ concentration of carbon vacancies,
the electron-hole puddles are created by the
superposition of contributions of randomly distributed
Gaussian impurities \cite{bardarson2007one} with a total number
of $N_{imp}$ = 16, i.e., the impurity concentration $n_{imp}$ = 1.1 $\times$ 10$^{13}$ cm$^{-2}$.
Gaussian potential at a position $\textbf{r$_{n}$}$ can
be written as follows: $V_{i} = \sum_{n=1}^{N_{imp}}\Delta_{n}\text{exp}\bm{\left[}-\bm{\left|}\textbf{r$_{i}$}-\textbf{r$_{n}$}\bm{\right|}^2\bm{/}\bm{\left(}2\upxi^{2}\bm{\right)}\bm{\right]}$,
where $\Delta$ is the impurity strength, and the impurity correlation length is taken to be $\xi$
= 10\emph{a} (\emph{a} = 0.142 nm is the \emph{C-C} distance) \cite{zhang2009origin}. Half of these impurities are chosen as positive and the other half as negative with the help of $\Delta$, which randomly fluctuates
within three different intervals: (i) $|\Delta|$ $<$ 0.1\emph{t}, (ii) $|\Delta|$ $<$ 0.3\emph{t}, and (iii) $|\Delta|$ $<$ 0.5\emph{t}.

Local density of states (LDOS) \cite{shytov2007atomic} is experimentally accessible
through a scanning tunneling microscope (STM) \cite{wang2013observing}
and is calculated by \(N(E,r)\) = $\sum_{i}|\Psi_{i}(r)|^{2}\delta(E-E_{i})$,
where $\Psi(r)$ is the normalized wave function, the energy \(E\) is identical to
applied bias voltage in STM measurements, and $E_{i}$ is the eigenenergy of the \emph{i}th state.
The LDOS is the spatially resolved density of states (DOS), which is calculated by summing
the discrete energy levels of the GQDs at a set of radial distances from the impurity,
ranging from r = 0.142 nm up to r = 1.136 nm. The summations are performed
by using a Gaussian membership function with a standard deviation of $\sigma$ = 0.2 eV
in a linearly spaced energy interval \emph{E} $\in$ [-2.5,2.5].
Since the effects of random disorders may differ from atom to atom,
these calculations are separately carried out for
each individual atom at the predefined radial distances, and
this is repeated in ten random disorder distributions for each of the above configurations \cite{text}. Finally,
the LDOS spectra per lattice site at various distances are extracted
by averaging over these samples.

It can be useful to discuss the effect of the vacuum size from
a different perspective before proceeding to the disordered cases.
The pristine hexagonal GQDs that differ in size are created, and
the discrete energy levels of them are summed over at the
impurity site r = 0.142 nm, as described above. Although such
a sum corresponding to a family of QBS is not necessary for the perfect vacuums,
it will provide a considerable advantage in the following sections.
All supercritical states are sequentially arranged within this family, which
contains the 1S$_{1/2}$ state as the first component \cite{shytov2007atomic}. Atomic collapse occurs
when this sharp peak in the electronic LDOS crosses just below the Dirac point (DP) \cite{wang2013observing},
which will be the energy origin in our calculations due to the formation of spatially extended resonances
at the negative energies \cite{van2017graphene}. Meanwhile,
the Fermi level (FL) moves down starting from the energy
origin as the coupling strength is increased within the half-filled model \cite{van2017graphene,polat2020collapse}.
To avoid too cumbersome notation, the critical coupling constant of the families of QBS is represented
by $\tilde{\beta_{c}}$, and only the response
of the spin-up Dirac fermions is studied for the perfect vacuums due to the presence of
a spin-independent central potential. The spin-up QBS families at the impurity site are shown in Fig. \ref{fig:fig_1}(a)
for the perfect GQDs consisting of 2814,
5514, and 10 806 carbon atoms. All families are pinned just below the DP
at $\tilde{\beta_{c}}$ = 1.0, revealing that
the effect of the Coulomb impurity is the same for all these GQDs, and the critical
bare valence charge is calculated to be $\tilde{Z_{c}}$ $\approx$ 2.73 by taking $\kappa$ = 6.

When point defects are evenly distributed between the sublattices, i.e., A (50$\%$) and B (50$\%$),
the FL continues to stay at the energy origin in the absence of the impurity \cite{pereira2006disorder,pereira2008modeling,kul2020electronic}
as shown in the inset of Fig. \ref{fig:fig_1}(c). In fact,
the FL is the same for both the perfect and disordered cases that validates the previous discussion on the
DP and the FL in our defect configurations when $\tilde{\beta}$ is turned on.
As for the spin symmetry, it is naturally broken in the disordered lattices \cite{altintacs2018defect}.
However, there is no difference between the spin-up
and spin-down families near $\tilde{\beta_{c}}$ as shown in Fig. \ref{fig:fig_1}(b) and Fig. \ref{fig:fig_1}(c), respectively.
As is clear from these two figures, all QBS families at the impurity site retreat from the DP
depending on the concentration of these defects, which are randomly distributed in the GQD
lattice consisting of 5514 atoms in the pristine case. Fig. \ref{fig:fig_1}(b) and Fig. \ref{fig:fig_1}(c) point out that
$\tilde{\beta}$ = 1.0 is no longer a critical
coupling constant, and it is
the first effect of point defects on atomic collapse.

\begin{figure}[t]
\begin{center}
\includegraphics[width=\linewidth, height=\textheight,keepaspectratio]{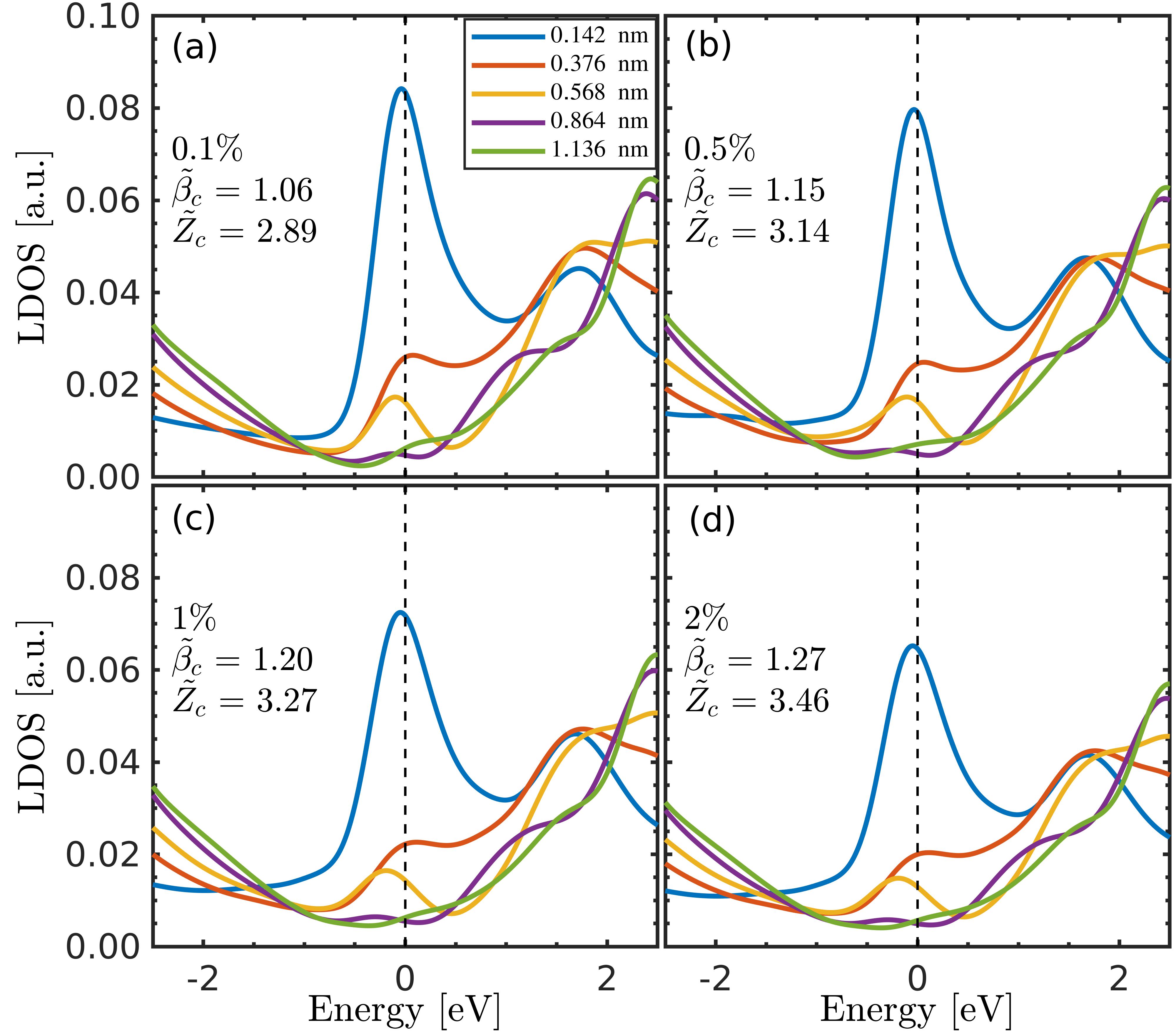}
\caption{\label{fig:fig_2} Defect-induced increase in the critical coupling constant $\tilde{\beta_{c}}$ for the concentrations of 0.1$\%$ in (a), 0.5$\%$ in (b), 1$\%$ in (c), and 2$\%$ in (d). Here spin-up and spin-down QBS families exactly overlap, and the different colored lines represent the corresponding radial distances from the impurity.}
  \end{center}
\end{figure}

These families transit from above to the edge of the DP
at different $\tilde{\beta_{c}}$ which is evident in Fig. \ref{fig:fig_2}(a)-(d).
The critical coupling constant gradually increases in proportion to the defect densities
and reaches $\tilde{\beta_{c}}$ = 1.27 for random dilution at
2$\%$ [see Fig. \ref{fig:fig_2}(d)]. Actually, these defects are ubiquitous
in the crystal structure \cite{eckmann2012probing}. For example, the Raman spectrum has
$\sim$ 0.5 G–to–2D intensity ratio for the high-quality graphene monolayer grown by chemical vapor
deposition (CVD) \cite{li2009large}, and this ratio indicates that
there is a finite defect density in graphene. As is clear from our numerical results, these structural peculiarities
can cause an increase in the critical threshold. On the other hand, the spectral shapes of all QBS families
are the same as of the defect-free case, especially in the vicinity of the impurity.
It can be inferred that atomic collapse can be similarly observed in the
imperfect lattices with the help of a higher valance charge.

\begin{figure}[b]
\begin{center}
\includegraphics[width=\linewidth, height=\textheight,keepaspectratio]{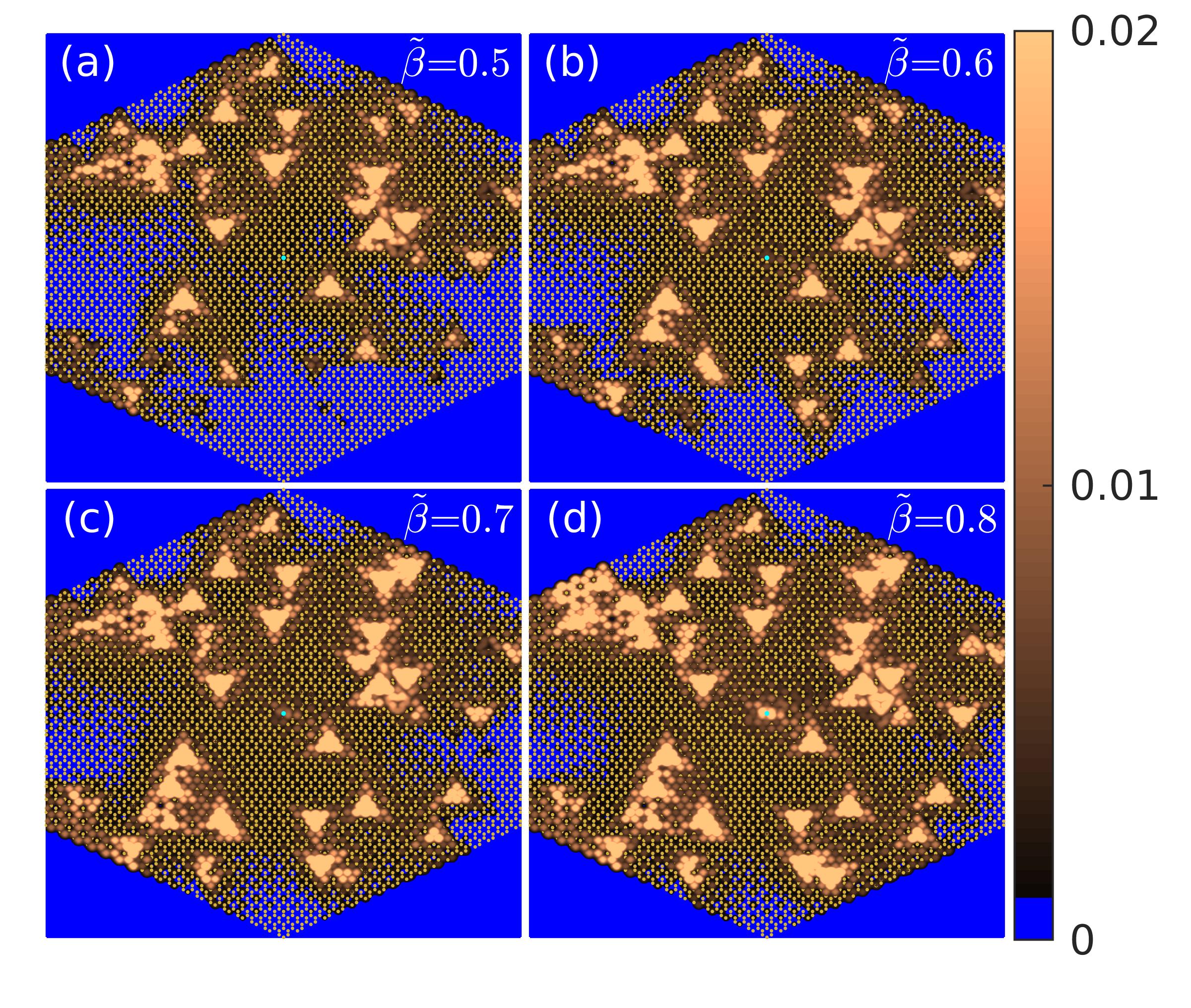}
\caption{\label{fig:fig_3}Response of the empty defect states below the DP to the Coulomb field for a representative sample with 1$\%$ defect density. Their spatial distributions are shown in (a)-(d) for $\tilde{\beta}$ = 0.5, 0.6, 0.7, and 0.8,
respectively. Upward (downward) triangular shapes belong to the unoccupied spin-up (spin-down) vacancy-induced states.
As is clear from (d), the 1S$_{1/2}$ state is formed at the center of QD marked by green dots.}
  \end{center}
\end{figure}

In the half-filled Hubbard model, the lowest energy states in the conduction band
are unoccupied vacancy-induced states whose energies are between 0 eV$<$\(E\)$<$0.4 eV
for 1$\%$ defect concentration [see the global DOS in the inset of Fig. \ref{fig:fig_1}(c)].
As $\tilde{\beta}$ is increased, these states successively dive into the negative energies.
However, there is no explicit crossing from the higher energy conduction states within the energy spectrums.
Therefore, of particular interest are these merging states below the DP, and the total probability density of them is calculated by p(r) =
$(1/2)\left[\sum_{E < 0}|\Psi(r)|^{2}-\sum_{E < E_{F}}|\Psi(r)|^{2}\right]$
in which both spin components are included. For a representative sample, p(r) is projected into the space at different coupling constants,
ranging from $\tilde{\beta}$ = 0.5 up to $\tilde{\beta}$ = 0.8.
Fig. \ref{fig:fig_3}(a)-(d) clearly show that whenever defect states dive just below the DP,
they are localized around the missing atoms by preserving their
characteristic triangular shapes and then demonstrate a striking
stability against the Coulomb impurity. On the other hand,
the weight of probability density around the impurity progressively increases,
but there is no formation of the first supercritical state at
$\tilde{\beta}$ = 0.5 nor at $\tilde{\beta}$ = 0.6, which are the critical
coupling constants of the 1S$_{1/2}$ state for the non-interacting \cite{shytov2007atomic} and interacting \cite{polat2020collapse} fermions in a clean vacuum, respectively. Finally, the 1S$_{1/2}$ state \cite{textt} comes into appearance
at $\tilde{\beta}$ = 0.8, despite not being a direct contribution of the bulk states to p(r).
Such a formation of the 1S$_{1/2}$ state is presumably due only to
the hybridized components of the diving defect states, and
the defect-induced increases in Fig. \ref{fig:fig_2}(a)-(d) actually originate
from the formation mechanism of the 1S$_{1/2}$ state.

\begin{figure}[b]
\begin{center}
\includegraphics[width=\linewidth, height=\textheight,keepaspectratio]{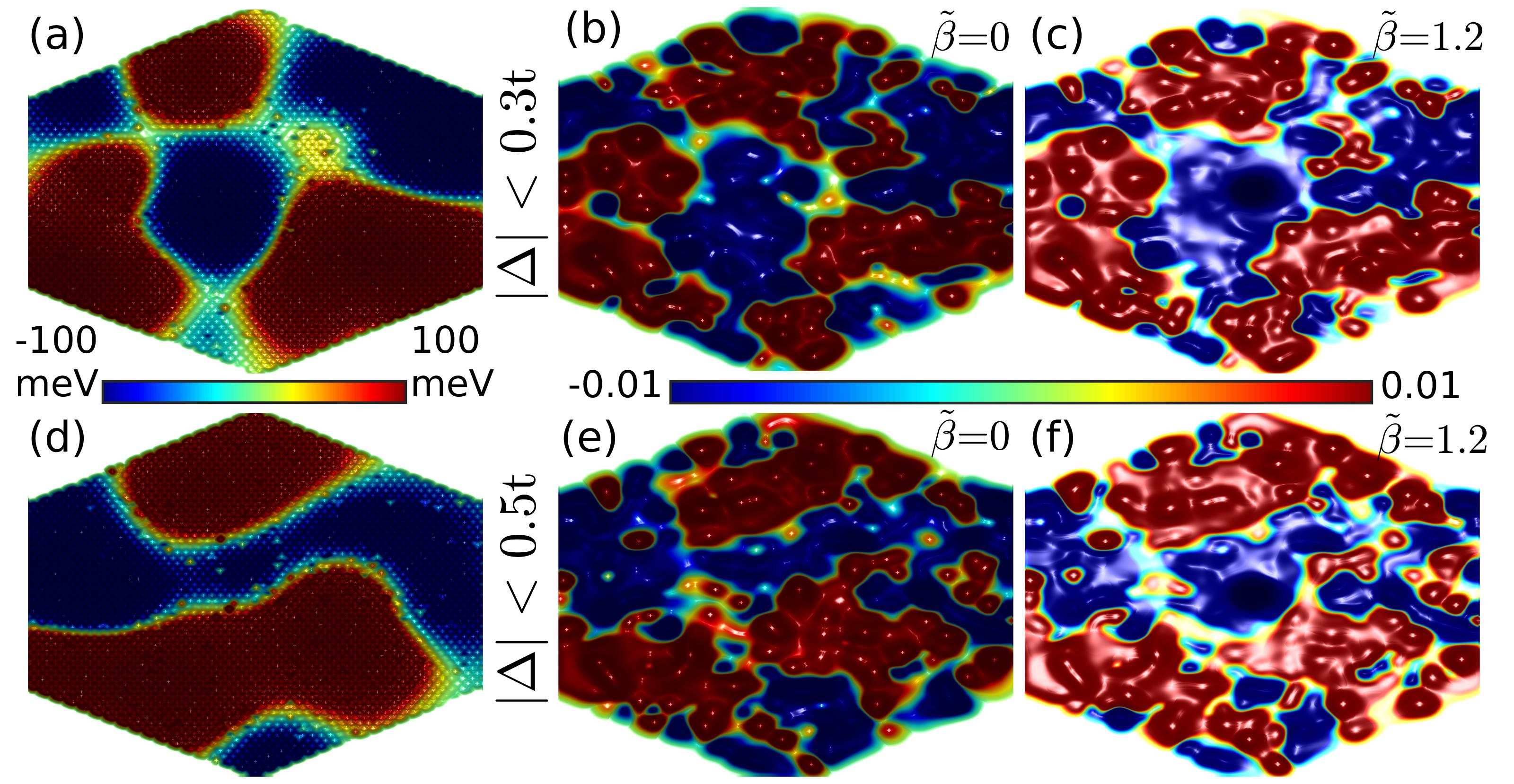}
\caption{\label{fig:fig_4}Upper panel: (a) averaged potential fluctuations for $|\Delta|$ $<$ 0.3\emph{t} (only $|\Delta|$'s averaged),
(b) the total electron-hole puddles accordingly are formed at $\tilde{\beta}$ = 0, and
(c) the reformation of these charge puddles at $\tilde{\beta}$ = 1.2. Lower panel: the same as the upper panel but now for $|\Delta|$ $<$ 0.5\emph{t}.}
  \end{center}
\end{figure}

Prior to the collapse experiments \cite{wang2012mapping,wang2013observing,mao2016realization,wong2017spatially},
monolayer graphene is grown by CVD
and then is transferred onto a hBN flake placed on
a SiO$_{2}$/Si substrate. To model the spatial potential fluctuations caused by such a substrate,
we randomly distributed Gaussian impurities for the set of vacuum disordered by
1$\%$ concentration of carbon vacancies. The averaged potential landscapes of
$|\Delta|$ $<$ 0.3\emph{t} and $|\Delta|$ $<$ 0.5\emph{t} are shown in
Fig. \ref{fig:fig_4}(a) and Fig. \ref{fig:fig_4}(d), respectively.
The resulting electron-hole puddles of both spin components show that
the electron puddles (red) appear in the positive potential regions, whereas the hole puddles (blue) manifest themselves
in the negative potential regions as can be seen
in Fig. \ref{fig:fig_4}(b) for $|\Delta|$ $<$ 0.3\emph{t} and Fig. \ref{fig:fig_4}(e) for $|\Delta|$ $<$ 0.5\emph{t}.
As $\tilde{\beta}$ is turned on, the charge inhomogeneities rearrange
themselves under the effect of the Coulomb potential. For example, at $\tilde{\beta}$ = 1.2, the electron-hole puddles of
$|\Delta|$ $<$ 0.3\emph{t} and those of $|\Delta|$ $<$ 0.5\emph{t} are mapped in Fig. \ref{fig:fig_4}(c) and
Fig. \ref{fig:fig_4}(f), respectively. Even if there is no significant change in the positions of the hole puddles
formed at the distances away from the center, those close to the center leave their positions and are centered
around the stronger Coulomb impurity. As will be seen below, such a reformation has
a significant effect on the critical threshold.

\begin{figure}[htb]
\begin{center}
\includegraphics[width=\linewidth, height=1.7in]{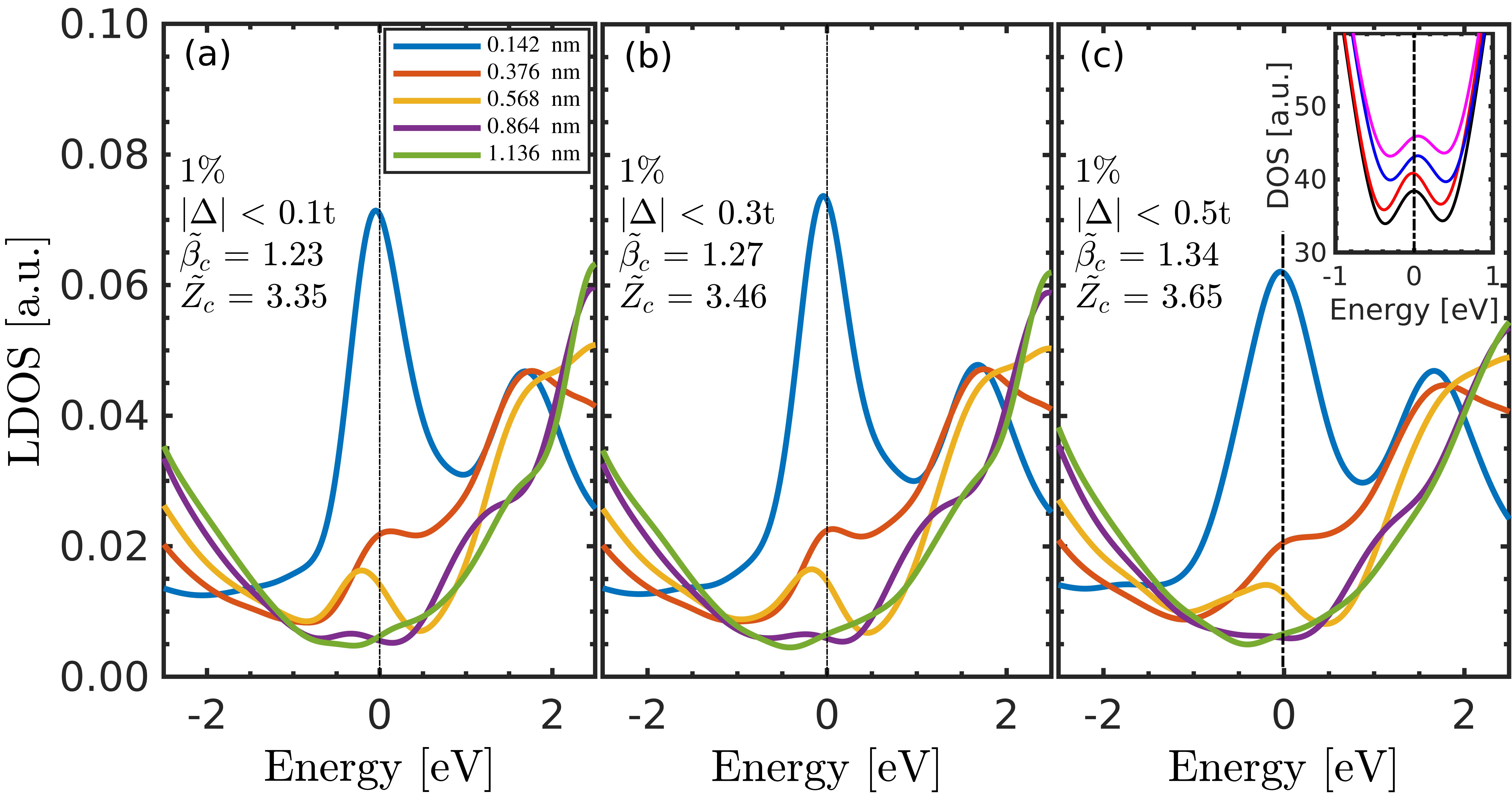}
\caption{\label{fig:fig_5} The effect of electron-hole puddles on the critical
threshold in (a)-(c), which are the same for both spin components. The inset in (c) shows averaged total DOS at
$\tilde{\beta}$ = 0, where black, red, blue, and purple lines represent $\Delta$ = 0, $|\Delta|$ $<$ 0.1\emph{t}, $|\Delta|$ $<$ 0.3\emph{t}, and $|\Delta|$ $<$ 0.5\emph{t}, respectively. For the sake of simplicity, a space between these lines is intentionally added.}
  \end{center}
\end{figure}

LDOS spectra in Fig. \ref{fig:fig_5}(a)-(c)
are calculated for the spin-up QBS family at the corresponding radial distances,
starting from the impurity site. When the positive and negative Gaussian impurities are distributed evenly,
the total DOS of the spin-up fermions at $\tilde{\beta}$ = 0 clearly reveal that the FL
is again around the energy origin for these configurations; see the inset in Fig. \ref{fig:fig_5}(c).
There is no significant shift in the minimum energy point at $\tilde{\beta}$ = 0, allowing us to take the energy origin
as the DP for the non-zero values of $\tilde{\beta}$. Similar to the previous cases,
whenever the sharp peak enters the negative energy spectrum, then atomic collapse has occurred.
The addition of Gaussian impurities causes to an increase in the critical threshold from $\tilde{\beta_{c}}$ = 1.20 [Fig. \ref{fig:fig_2}(c)] up to $\tilde{\beta_{c}}$ = 1.34 [Fig. \ref{fig:fig_5}(c)], and the critical valance charge
is estimated to be as high as $\tilde{Z_{c}}$ = 3.65. In addition, we also study the point defect-free GQD consisting of 5514 atoms for ten random distributions of $|\Delta|$ $<$ 0.5\emph{t}, and the critical threshold reaches to $\tilde{\beta_{c}}$ = 1.10 (not shown here), which is $\tilde{\beta_{c}}$ = 1.0 [Fig. \ref{fig:fig_1}(a)] in its clean case. It can be noted that the increments in the critical threshold are independent of the sign of the substrate-induced potential where the Coulomb impurity is placed
and are directly proportional to the strengths of Gaussian impurities. As a result, $\tilde{\beta_{c}}$
seems to be highly influenced by the disorders within the vacuum itself.

In bulk graphene, a series of LDOS measurements performed by a STM reveals that a cluster,
composed of four calcium dimers in the charge state of +1$|e|$, is needed to form an
infinite family of QBS at just above the DP [see Fig. 1(D)
in Ref. \cite{wang2013observing}]. Therefore, the critical
bare valance charge should be slightly greater than $\tilde{Z_{c}}$ $\gtrsim$ 4
in the experiment. Accordingly, the calculated values of $\tilde{Z_{c}}$ are approaching to
that of the experiment, and adding these experimentally
relevant factors to the Coulomb impurity problem opens a new route
towards such experimental results \cite{wang2012mapping,wang2013observing}.
These findings can be useful in interpreting the experimental
results of positively charged Coulomb impurities, even if they
exceed the theoretical critical value. Results of this paper can be tested via Ar$^{+}$ ion bombarded \cite{lucchese2010quantifying},
He$^{+}$ ion irradiated \cite{chen2009defect}, and hydrogenated \cite{bostwick2009quasiparticle} graphene.
The latter can be achieved by transferring
CVD graphene samples at different H coverages \cite{bostwick2009quasiparticle} onto a hBN/SiO$_{2}$/Si device,
which facilities to control bias and back-gate voltages. Impurities such as cobalt trimmers \cite{wang2012mapping} and calcium dimers \cite{wang2013observing,wong2017spatially} can be gathered in a defect-rich region by atomic manipulation of
them with the help of STM, and an artificial supercritical atom can be created
from these subcritical impurities. Once the DP has been determined,
LDOS spectra can be measured at different radial or lateral distances.
There should be an increase in the critical threshold due to the partial removal of
the $\pi_{z}$ states depending on the concentration of H.
\begin{acknowledgments}
This work was supported by
The Scientific and Technological Research Council of Turkey
(TUBITAK) under 1001 Grant Project No. 116F152.
\end{acknowledgments}


\begin{thebibliography}{38}
\bibitem{text}
Consider the case with $\%1$ defect concentration,
we have ten different samples, each with totally random distributions.
For each of these samples, we randomly distribute Gaussian
impurities with $|\Delta|$ $<$ 0.1\emph{t}. This is also performed for
$|\Delta|$ $<$ 0.3\emph{t} and $|\Delta|$ $<$ 0.5\emph{t} in
the presence of $\%1$ defect concentration.
\bibitem{textt}
Its shape is exactly the same as the 1S$_{1/2}$ state of
the defect-free GQDs (not shown here), and
this formation is observed for all samples at
around $\tilde{\beta}$ = 0.8 for random dilution at
1$\%$.
\end{thebibliography}

\providecommand{\noopsort}[1]{}\providecommand{\singleletter}[1]{#1}%
\end{document}